\documentclass[fleqn,12pt,twoside]{article}
\usepackage[headings]{espcrc1}

\readRCS
$Id: espcrc1.tex,v 1.2 2004/02/24 11:22:11 spepping Exp $
\usepackage{graphicx}

\title{
What can we learn from hydrodynamic analysis of elliptic flow?
        \thanks{
	This work was supported in part by the United States
	Department of Energy under Grant No.~DE-FG02-93ER40764.
	}
}

\author{Tetsufumi Hirano\address{Department of Physics,
	Columbia University, New York, NY 10027, USA}
}

\runtitle{
What can we learn from hydrodynamic analysis of elliptic flow?
}
\runauthor{
Tetsufumi Hirano
}

\begin{document}

\maketitle

\begin{abstract}
We can establish a new picture, 
the perfect fluid 
strongly coupled quark gluon plasma
(sQGP) core and
the dissipative hadronic corona,
of the space-time evolution of
produced matter in relativistic heavy ion collisions
at RHIC.
It is also shown that the picture works well
also in the forward rapidity region through an analysis
based on a new class of the hydro-kinetic model
and is a manifestation of deconfinement.
\end{abstract}

\section{INTRODUCTION}

Recently, physicists at Brookhaven National Laboratory
made an announcement 
that ``RHIC serves the perfect liquid" \cite{BNL}.
The agreement of hydrodynamic predictions
\cite{Huovinen:2003fa}
of integrated
and differential elliptic flow and radial flow patterns
with Au+Au data at RHIC energies
\cite{Ackermann:2000tr,Adcox:2002ms,Back:2002gz,Ito} 
is one of the main lines of the announcement.
We first study the sensitivity of this conclusion
to different hydrodynamic assumptions 
in the hadron phase.
It is found that an assumption of chemical equilibrium
with neglecting viscosity in the hadron phase
in hydrodynamic simulations causes accidental reproduction
of transverse momentum spectra and 
differential elliptic flow data.
From a systematic comparison of hydrodynamic
results with the experimental data, dissipative effects
are found to be mandatory in the hadron phase.
Therefore, what is discovered at RHIC is not only
the perfect fluidity of the strongly coupled quark gluon plasma
(sQGP) core
but also its dissipative hadronic corona.
Along the lines of these studies,
we develop a hybrid dynamical model
in which a \textit{fully three-dimensional}
hydrodynamic description of the QGP phase
is followed by a kinetic description of the hadron phase 
\cite{nonaka}.
We show rapidity dependence of elliptic flow from this hybrid model
supports the above picture.
Finally, we argue that this picture is a manifestation of
deconfinement transition, namely,
a rapid increase of entropy density
in the vicinity of the QCD critical temperature
as lattice QCD simulations have been predicted.

\section{SQGP CORE AND DISSIPATIVE HADRONIC CORONA}

A perfect fluid in the QGP phase 
and the first order phase transition to the hadron phase
is assumed 
in most hydrodynamic simulations \cite{hama}.
While one can find various assumptions in the hadron phase,
e.g.~(1) ideal and chemical equilibrium (CE) fluid,
(2) ideal and chemically frozen fluid 
(or partial chemical equilibrium, PCE), or
(3) non-equilibrium resonance gas via hadronic cascade models (HC).
Hydrodynamic results are compared with the current
differential elliptic flow data, $v_2(p_T)$,
in Fig.~20 in Ref.~\cite{Adcox:2004mh}
with putting an emphasis on the difference of
assumptions in the hadron phase.
The classes CE and HC reproduce
the pion data well.
On the contrary, results from the second class, PCE, 
deviate from these hydrodynamic results and
experimental data.
In order to claim the discovery of perfect fluidity
from the agreement of hydrodynamic results 
with $v_2(p_T)$ data,
we need to understand the difference among hydrodynamic
results and the deviation from data.
$v_2$ is roughly proportional to $p_T$
in low $p_T$ region for pions.
In such a case, the slope of $v_2(p_T)$
can be approximated by
$v_2/\langle p_T \rangle$.
Integrated $v_2$ is generated in the early
stage of collisions. Whereas \textit{differential}
$v_2$ can be sensitive to the late hadronic stage
since $dv_2(p_T)/dp_T \approx v_2/\langle p_T \rangle$
indicates interplay between elliptic flow and radial flow.
\begin{figure}[htb]
\begin{minipage}[t]{80mm}
\includegraphics[width=75mm]{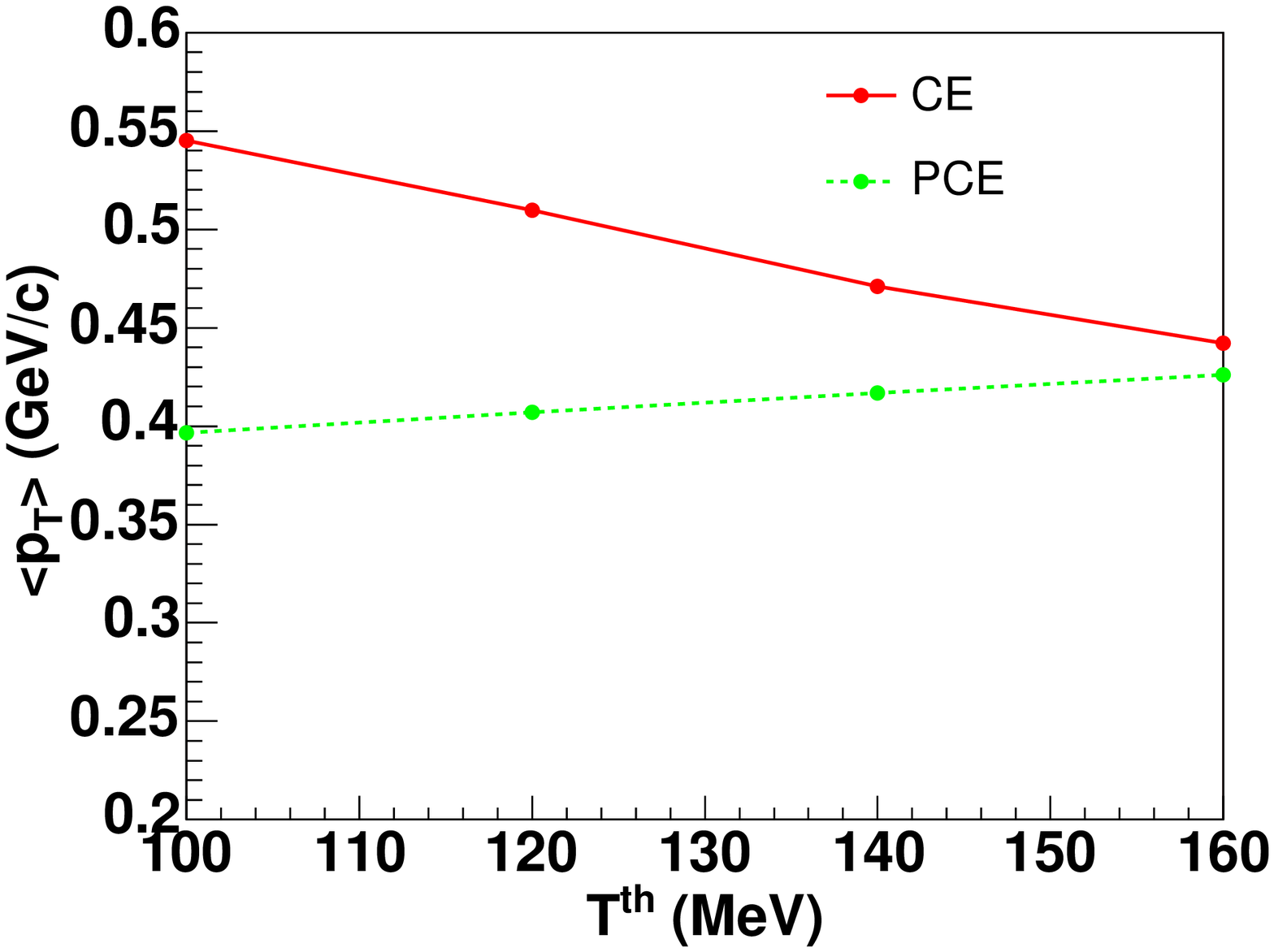}
\caption{$\langle p_T \rangle$ for
pions as a function of thermal
freezeout temperature at $b=5$ fm at RHIC.
Solid (dashed) line shows a result with
an assumption of an ideal, chemical equilibrium (chemically frozen)
hadronic fluid.
}
\label{fig:meanpt}
\end{minipage}
\hspace{\fill}
\begin{minipage}[t]{75mm}
\includegraphics[width=74mm]{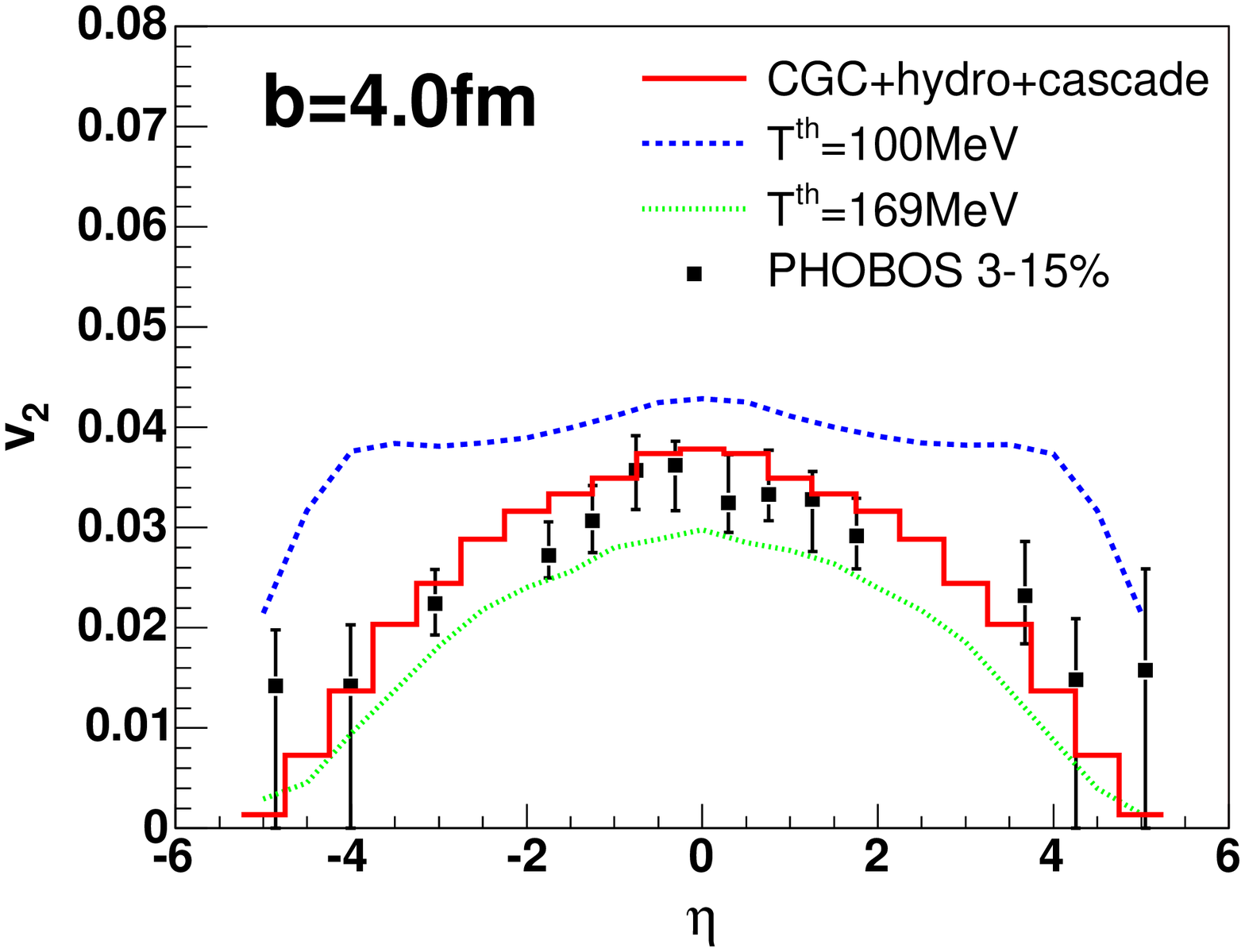}
\caption{
Pseudorapidity dependence of elliptic flow
from hydro and hydro+hadronic cascade models are
compared with data \cite{Back:2002gz}.
}
\label{fig:v2eta}
\end{minipage}
\end{figure}
In Fig.~\ref{fig:meanpt},
thermal freezeout temperature $T^{\mathrm{th}}$ dependence of
$\langle p_T \rangle$ for pions including
contribution from resonance decays
are calculated at a RHIC energy by utilizing
hydrodynamic simulations \cite{hg}.
$\langle p_T \rangle$ for pions 
in the chemically frozen hadronic fluid
decreases with decreasing $T^{\mathrm{th}}$.
This is due to longitudinal $pdV$ work done by fluid elements.
Whereas $\langle p_T \rangle$ in the chemical equilibrium case
increases during expansion.
The total number of particles in a fluid element
decreases due to the assumptions of
 chemical equilibrium and entropy conservation in the
hadron phase.
Then the total energy is distributed among the smaller number
of particles as a fluid element expands.
These are the reasons why the different behavior 
of $\langle p_T \rangle$ appears
according to the assumption of chemical equilibrium/freezeout.
Under the chemical equilibrium assumption
in ideal hydrodynamic simulations,
increasing $\langle p_T \rangle$ is
commonly utilized so far to fix  
$T^{\mathrm{th}}$
by fitting $p_T$ slope.
However, this is attained only
by neglecting data of particle ratio.
If particle ratios are fixed
properly in hydrodynamic simulations
to reproduce the data,
$p_T$ slopes, especially for protons, are hardly reproduced.
The same is true for differential elliptic flow:
$dv_2(p_T)/p_T \approx v_2/\langle p_T \rangle$
is reproduced by canceling increase behaviors of 
both $v_2$ and $\langle p_T \rangle$
under chemical equilibrium assumption \cite{hg}.
Agreement of the results from the CE model
with $p_T$ spectra and $v_2(p_T)$ data
is merely an accident in the sense that this model
makes full use of neglecting particle ratio
to reproduce them.
So the HC model turns out to be the only
model which is able to reproduce
particle ratio, $p_T$ spectra and $v_2(p_T)$ data.
Therefore a picture of 
the dissipative hadronic corona
together with
the perfect fluid sQGP core 
is consistent with these
experimental data observed at RHIC.

\section{3D HYDRO AND HADRONIC CASCADE MODEL}
According to the discussion
in the previous section,
we incorporate a hadronic cascade model, JAM \cite{jam},
into our previous framework,
the ``CGC+hydro+jet" model \cite{hiranonara}.
Figure 2 shows pseudorapidity dependences of $v_2$
from this hybrid model 
and ideal 3D hydrodynamics with $T^{\mathrm{th}}=100$ and 169 MeV.
Here critical temperature
and chemical freezeout temperature
are taken as being 
$T_c = T^{\mathrm{ch}}=170$ MeV
in the hydrodynamic model.
In the hybrid model,
the switching temperature
from a hydrodynamic description to a kinetic one
is taken as $T_{\mathrm{sw}}=169$ MeV.
Ideal hydrodynamics with $T^{\mathrm{th}}=100$ MeV which is so chosen
to generate enough radial flow gives
a trapezoidal shape of $v_2(\eta)$ \cite{hirano3d}.
A large deviation between data \cite{Back:2002gz}
and the ideal hydrodynamic result
is seen especially in forward/backward rapidity regions.
When hadronic rescattering effects are taken through
the hadronic cascade model
instead of perfect fluid description of the hadron phase,
$v_2$ is not so generated in the forward region due to the
dissipation and, eventually, is consistent with
the data. 
So the perfect fluid sQGP core and the dissipative hadronic
corona picture works well also in the forward region.

\section{SUMMARY: WHAT HAVE WE LEARNED?}

We can establish a new picture of space-time evolution
of produced matter from 
a careful comparison of hydrodynamic
results with experimental data
observed at RHIC.
What is the physics behind this picture?
$\eta/s$ is known to be a good measure to see
the effect of viscosity where $\eta$
is the shear viscosity and $s$ is the entropy density.
Figures 3 and 4 show possible scenarios
for temperature dependence
of $\eta$ and $\eta/s$ deduced from the
discussion in the previous sections.
$\eta/s$ is small in the QGP phase,
which might be comparable 
with the minimum value $1/4\pi$ \cite{Son}, 
and the perfect fluid assumption
can be valid. While $\eta/s$ becomes huge
in the hadron phase and the dissipation
cannot be neglected.
Shear viscosities of both phases 
are found to give $\eta \sim 0.1$ GeV/fm$^2$
around $T_c$ \cite{hg}.
So shear viscosity itself appears to increase with temperature
monotonically.
The ``perfect fluid'' property of the sQGP is thus not
due to a sudden reduction of the viscosity
at the critical temperature $T_c$, but to
a sudden increase of the entropy density 
of QCD matter and is therefore an
important signature of deconfinement.
\begin{figure}[htb]
\begin{minipage}[t]{80mm}
\includegraphics[width=75mm]{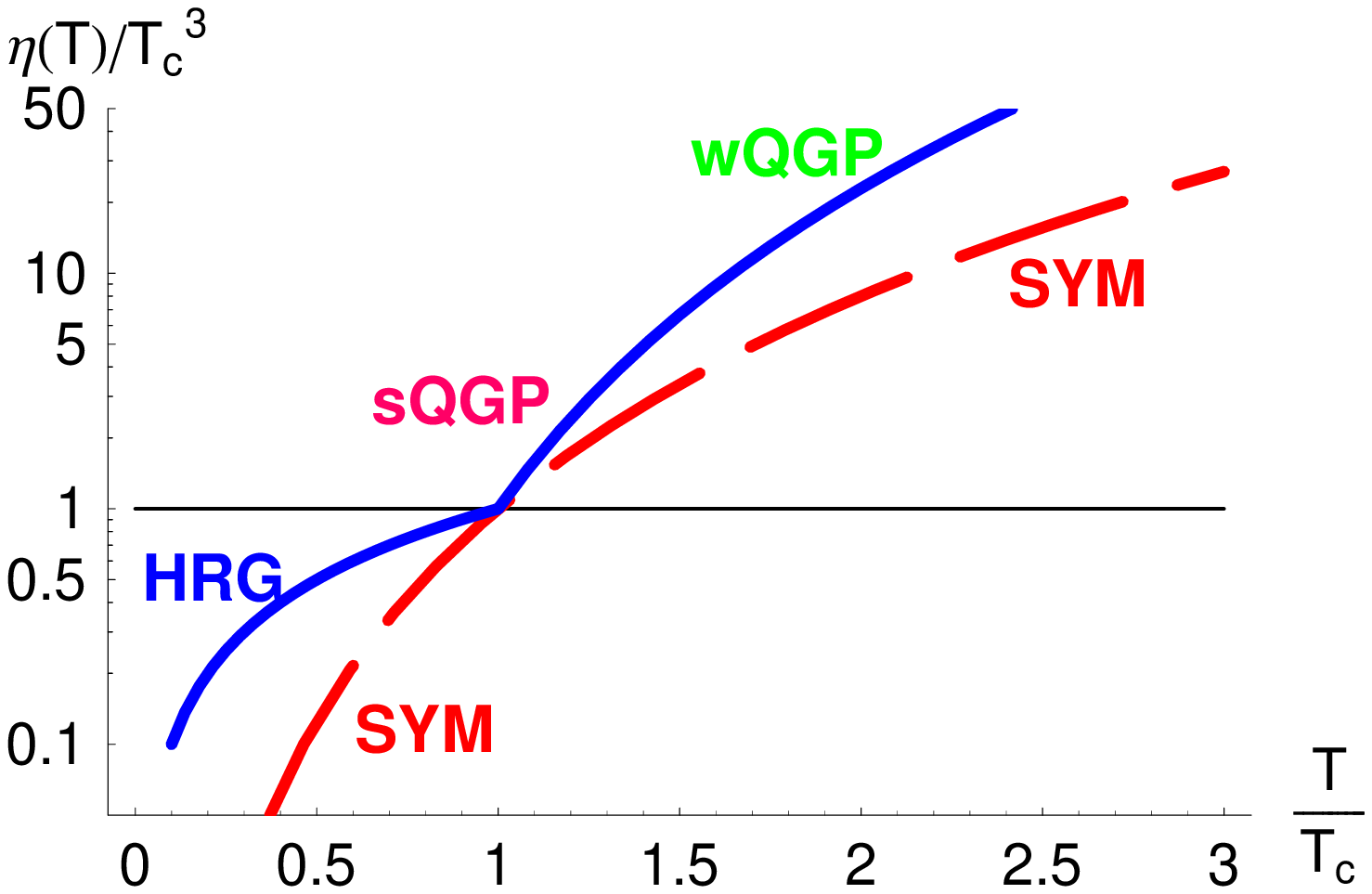}
\caption{
Illustration of
the shear viscosity as a function of temperature.
SYM, HRG, and wQGP represent, respectively, supersymmetric Yang-Mills model,
hadronic resonance gas, and weakly coupled QGP.
}
\label{fig:eta}
\end{minipage}
\hspace{\fill}
\begin{minipage}[t]{75mm}
\includegraphics[width=74mm]{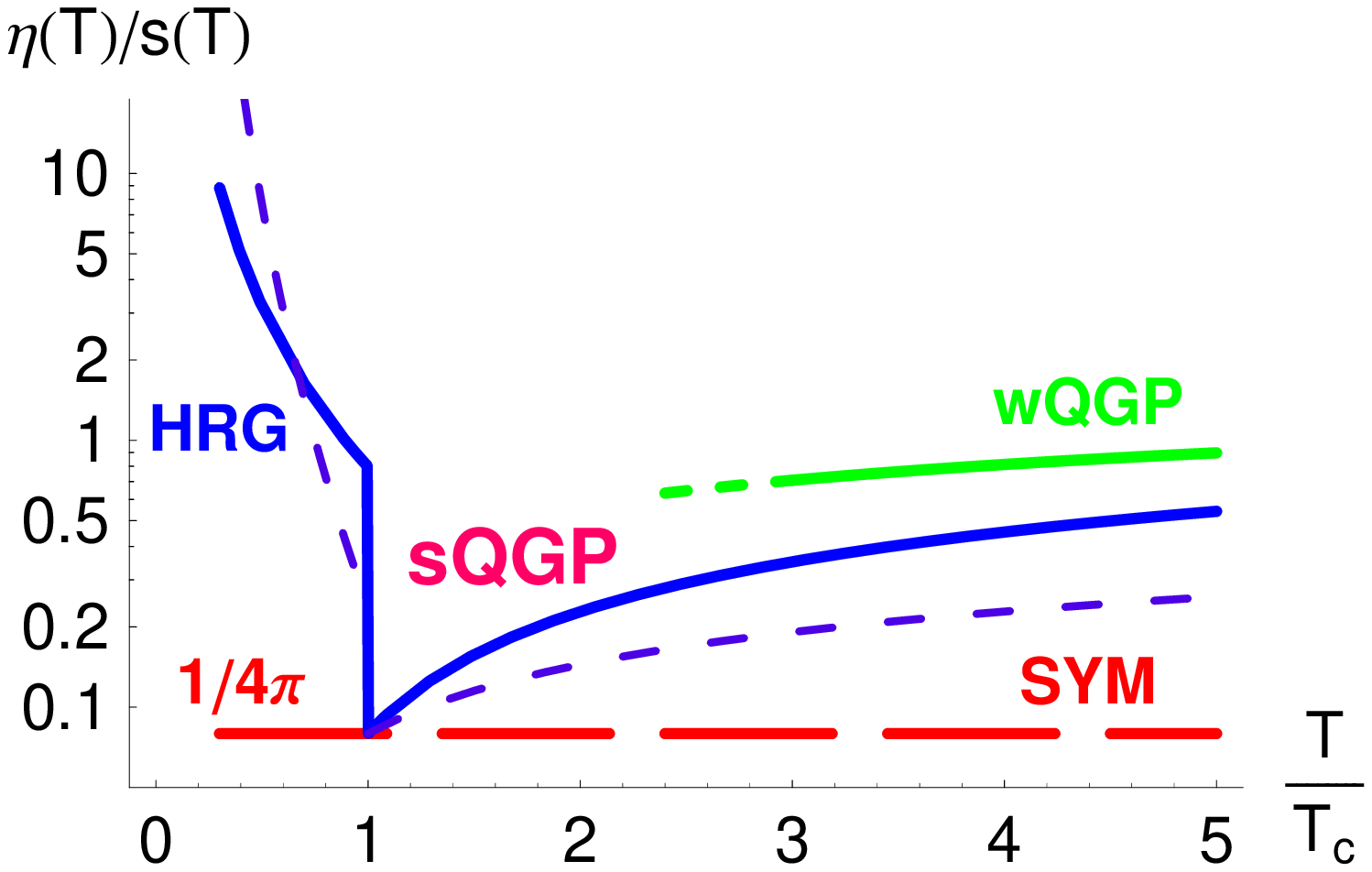}
\caption{
Illustration of the rapid variation of the 
\textit{dimensionless ratio}
of the shear viscosity, $\eta(T)$, to the entropy density, $s(T)$.
}
\label{fig:etaovers}
\end{minipage}
\end{figure}

\vspace{12pt}

The author would like to thank M.~Gyulassy and Y.~Nara
for collaboration and fruitful discussion.


\begin{thebibliography}{9}
\bibitem{nonaka} A study based on a similar approach 
(3D Lagrangean hydrodynamics + UrQMD) is also presented
in this conference: C.~Nonaka, these proceedings.
\end{thebibliography}
\end{document}